\documentclass[10pt,a4paper,twoside]{article}
\usepackage{epsfig}
\usepackage{baltlat6}
\usepackage{array}
\usepackage{here}
\begin{document}
\ \
\vspace{0.5mm}
\setcounter{page}{523}
\vspace{8mm}

\titlehead{Baltic Astronomy, vol.\,20, 523--530, 2011}

\titleb{COMPARISONS AND COMMENTS ON ELECTRON AND ION\\
IMPACT PROFILES OF SPECTRAL LINES}

\begin{authorl}
\authorb{Sylvie Sahal-Br\' echot}{1},
\authorb{Milan S. Dimitrijevi\' c}{2,1},
\authorb{Nebil Ben Nessib}{3}
\end{authorl}

\begin{addressl}
\addressb{1}{Paris Observatory, LERMA CNRS UMR 8112, UPMC,\\  5 Place Jules Janssen, 92190 Meudon, France;
sylvie.sahal-brechot@obspm.fr}
\addressb{2}{Astronomical Observatory, Volgina 7,\\  11060 Belgrade 38, Serbia; mdimitrijevic@aob.bg.ac.rs.}
\addressb{3}{INSAT (National Institute of Applied Sciences and Technology),\\ University of Carthage, Tunis, Tunisia; nebil.bennessib@planet.tn}
\end{addressl}

\submitb{Received: 2011 August 8; accepted: 2011 August 15}

\begin{summary} Stark broadening theory is currently operated for calculating widths and shifts of spectral lines that are needed for spectroscopic diagnostics and modelling in astrophysics,  laboratory and technological plasmas. We have calculated a  great number of data, obtained through the impact semi-classical perturbation theory: tables have been published for neutral atom and ion emitters, and typical temperatures, electron and ion densities. They are currently implemented in the STARK-B database which participates to the European effort  within the VAMDC (Virtual Atomic and Molecular data Centre). Despite of that, a great number of data are still missing and their orders of magnitude would at least be welcome. In the present paper, we will revisit  and compare the orders of magnitudes and trends of the impact Stark widths and shifts, by considering their semiclassical perturbation expressions. We will also provide fitting formulae which  are essential for the modelling codes of stellar atmospheres and envelopes.
\end{summary}

\begin{keywords} atomic data -- atomic processes -- line: profiles -- astronomical databases \end{keywords}

\resthead{Electron and ion impact spectral line profiles}
{S. Sahal-Br\'echot, M.S. Dimitrijevi\'c and N. Ben Nessib}

\sectionb{1}{INTRODUCTION}

Pressure broadening of spectral lines arises when an atom, ion, or molecule which emits or absorbs light in a gas or a plasma, is perturbed by its interactions with the other particles of the medium. In the present paper, we will consider atom or ion emitters and electron and ion colliders. It is the so-called Stark broadening. It has been extensively developed for about 50 years. It is now currently used for spectroscopic diagnostics and modelling.
In astrophysics, with the increasing sensitivity of observations and spectral resolution, in all domains of wavelengths from far UV to infrared, it has become possible to develop realistic models of interiors and atmospheres of stars and interpret their evolution and the creation of elements through nuclear reactions. This requires the knowledge of numerous profiles, especially for trace elements, which are used as useful probes for abundance determinations. For white dwarfs in particular, Stark broadening is the dominant line broadening process. Hence, calculations based on a simple but enough accurate and fast method, are useful for obtaining numerous results. Ab initio calculations are a growing domain of development. Nowadays, the access to such data via an on line database becomes crucial. This is the object of STARK-B (Sahal-Br\'echot et al. 2008), which is a collaborative project between the Paris Observatory and the Astronomical Observatory of Belgrade. It is a database of calculated widths and shifts of isolated lines of atoms and ions due to electron and ion collisions. It is devoted to modelling and spectroscopic diagnostics of stellar atmospheres and envelopes. In addition, it is relevant to laboratory plasmas, laser equipments and technological plasmas. It is a part of VAMDC (Virtual Atomic and Molecular Data Centre, Dubernet et al. 2010, Rixon et al. 2010), which is an European Union funded collaboration between groups involved in the generation and use of atomic and molecular data.

In the present paper, we will briefly recall the important points of the theory. Then we will revisit  the orders of magnitudes and trends of the widths and shifts in the impact approximation for electron and ion colliders.
This is useful for providing interpolation or extrapolation formulae  for missing data, since it is impossible to calculate all the data necessary to the modelling. We will also provide fitting formulae, which are essential for the modelling codes of stellar atmospheres and stellar envelopes. The coefficients of our proposed  fitting formulae will be implemented  in STARK-B.

\sectionb{2}{REMIND OF THE IMPACT LINE BROADENING FOR ISOLATED LINES  }

Stark broadening theory in the impact approximation is based on the founding papers  by Baranger (1958a,b,c). The impact approximation is the first basic one: $\rho$  being a typical impact parameter and $v$ the relative velocity, the duration of a collision or collision time $\tau = \rho/v$ must be much smaller than the mean interval between two collisions. which is of the order of the inverse of the line width $w$ (in angular frequency units). So the collisions between the radiating atom (or ion) act independently and are additive. It is quite always valid for electron collisions and is generally valid for collisions with positive ions in the conditions of stellar atmospheres (Sahal-Br\'echot 1969a,b).
The second basic approximation is the complete collision approximation: the radiating atom has no time to emit (or absorb) a photon during the collision process. In other words, the collision time $\tau$ must be very much smaller than the time interval between two emissions (or absorptions) of photons. The latter is of the order of the inverse of the detuning $\Delta \omega$. So, in the far wings, the atom can emit photons before the perturber has any time to move, and thus the process becomes quasistatic.
In the line center, the impact approximation and the complete collision approximation are together valid, and the line broadening theory becomes an application of the theory of collisions between the radiating atom and the surrounding perturbers.

Then we will limit our study to the case of isolated lines: the levels of the $i-f$ transition broadened by collisions do not overlap with the neighbouring perturbing levels which are likely to modify the broadening by introducing optical coherences. So we will consider  in the present paper neither hydrogen nor hydrogenic ionic lines, nor some specific helium lines and nor some lines arising from Rydberg levels.

This leads to a Lorentz  line profile characterized by a width $w$ (full half-width at half-maximum) and a shift $d$ which depend on the physical conditions of the medium (temperature $T$ and density $N$ of the perturbers). Owing to the impact approximation, $w$ and $d$ are proportional to the density. The width of the $i-f$ line can be expressed as a sum over the inelastic cross-sections  $\sigma_{ii'}(v)$  and $\sigma_{ff'}(v)$  ($i'$ and $f'$ are the so-called perturbing levels) and over an elastic contribution  $\sigma_{el}(v)$ that are integrated over the Maxwell distribution of velocities $f(v)$. The shift can be expressed in terms of another elastic contribution, cf. Baranger (1958c).

In addition, the Debye screening effect which can be important at high densities must be taken into account. This decreases $w$ and $d$ which are  thus not proportional to the density.

Finally, we will remark that the fine and hyperfine structure can be neglected during  collisions with electrons. This is due to the fact that the electron spin of the atom has no time to rotate (Larmor precession) during the collision time $\tau$, because the relative velocity atom-perturber $v$ is large. Consequently, but only in $LS$ coupling,  the widths and shifts due to electron collisions are equal for the different lines of a multiplet. Departures from $LS$ coupling can be important for heavy atoms or for highly charged ions and then the fine structure widths and shifts can be different. For collisions with ions, the relative velocity is smaller, and the preceding condition is  not completely fulfilled for the electronic spin, but the widths and shifts of the fine structure line are not very different. In fact the hyperfine structure is always negligible during the collisions. However, if the fine structure (or hyperfine structure) splitting is not negligible, the components must be added (taking into account their shift and their relative intensities) for obtaining the global profile. This is the case of the hyperfine Mn II lines (Popovi\'c et al. 2008) and of the fine hydrogen Balmer lines (Stehl\'e and Feautrier 1985).

\sectionb{3}{THE SEMICLASSICAL-PERTURBATION THEORY-SCP}

Most of our calculations have been performed with the semi-classical-perturbation method (SCP) developed by Sahal-Br\'echot (1969a,b) and further papers:  Sahal-Br\'echot (1974) for complex atoms, Fleurier et al. (1977) for inclusion of Feshbach resonances in elastic cross-sections of radiating ions, and by Mahmoudi et al. (2009) for very complex atoms. The numerical codes have been updated and operated by Dimitrijevi\'c and Sahal-Br\'echot  (1984) and then by many further papers. The accuracy is about 20\% for the widths but less for the shifts, due to oscillations in the integration over the impact parameters in the neighbourhood of the cut-off region. Cf. Sahal-Br\'echot (2010) for a brief remind of other methods.

We now focus on  the physical quantities  that enter the expressions of $w$ and $d$. This permits to understand their behaviours and trends  as functions of atomic structure (oscillator strengths $f_{ii'}$, $f_{ff'}$, energy levels $E_{i}$, $E_{i'}$, $E_{f}$, $E_{f'}$), temperature $T$, charges of the radiating ion $Z_A$ and perturber $Z_P$, reduced mass atom-perturber $\mu$. The incoming perturber moves along a straight path (neutral radiating atom) or an hyperbola  for a radiating ion (Sahal-Br\'echot 1969a,b). We refer to Sahal-Br\'echot (1969a,b) for all the required formulae which will not be recalled here. We will only  have a look on the results. We will focus on widths because shifts are often less accurate.

First, we consider the role of collision strengths.

We begin by lines of neutral atoms on the example of Mg I lines (Dimitrijevi\'c \& Sahal-Br\'echot 1996 and STARK-B). Figure 1 shows the width of Mg I $3s^{1}S-3p^{1}P^{o}$  as a function of the temperature $T$ in Kelvin, for an electronic density of $10^{12}$ cm$^{-3}$. The difference of energy between  the initial (or final) level and the closest perturbing level is $\Delta E_{\mathrm{min}} = 8451.64$ cm$^{-1}$, and $\Delta E_{\mathrm{min}}/kT=1.2$ at $5000$ K.
 Rather distant levels are involved, hence at those temperatures inelastic collisions are completely negligible for impact ions. Elastic collisions are mostly due to the quadrupolar potential (cf. right part of Figure 1), and since the quadrupolar contribution does not depend on the reduced mass, this explains why the width due to impact proton and the soup of ions (same charge as that of protons) are equal.
\begin{figure}[!tH]
\vbox{
\centerline{\psfig{figure=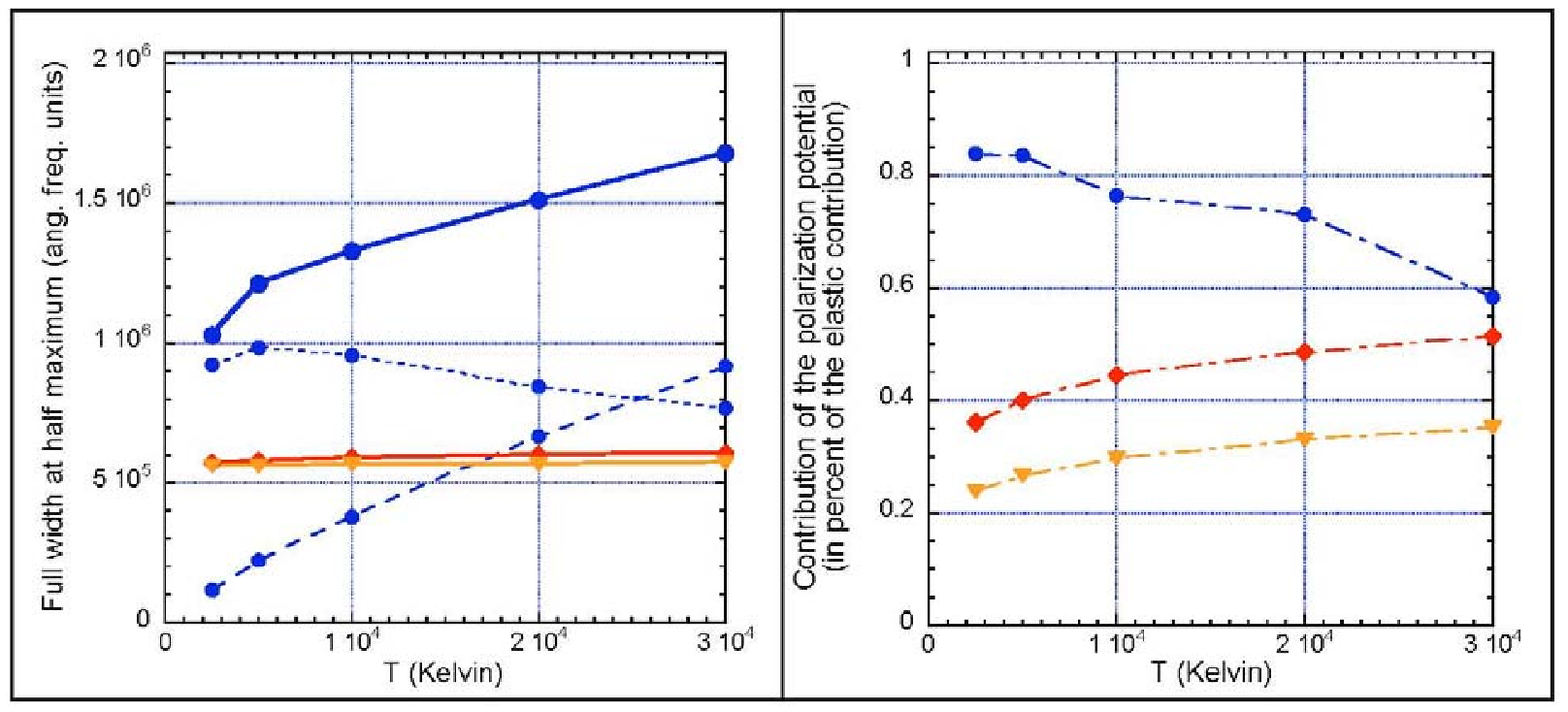,width=90mm,angle=0,clip=}}
\captionb{1}
{Mg I $3s^{1}S-3p^{1}P^{o}$ :
Left part: $w$ in angular frequency units as a function of the temperature $T$ in Kelvin, the electronic density is $10^{12}$ cm$^{-3}$. Full lines: total widths; dotted lines: elastic contributions; dashed lines: inelastic contributions;  circles: electrons; diamonds: protons; triangles: soup of ions.
Right part: contribution of the polarization  potential, in percent of the elastic contribution
}}
\end{figure}

Table 1  shows the increasing contribution of impact ions when higher levels are involved. It displays the ratio of the  width due to electron collisions to the width due to Fe II collisions as a function of $T$ for  Mg I $n+1, l+1- n, l$, with $l=n$. The width due to impact ions becomes higher than the one due to impact electrons.
\bigskip
{\smallbf\ \ Table 1.}{\small\
Mg I $n+1, l+1- n, l$, with $l=n$: ratio of the  width due to electron collisions to the width due to Fe II collisions as a function of $T$ (in $10^{3}$ Kelvin). $N=10^{10}$ cm$^{-3}$}.
\begin{tabbing}
$T$ ($10^{3}$ Kelvin) \=  $6h-5g$  \=  $7i-6h$  \=  $8j-7i$ \=  $9k-8j$  \=  $10l-9k$  \\
2.5  \>   5.08  \>  3.59  \>  2.30  \>  1.44    \>  0.72          \\
4.5  \>  4.35  \>  2.80  \>  1.71  \>   0.998  \> 0.474        \\
6.0  \>  3.97  \>  2.48  \>  1.45  \>   0.800  \>  0.40  \\
\end{tabbing}
Figure 2 shows the details of the contributions for two lines : Mg I $5g^{1}G-6h^{1}H^{o}$, where the closest perturbing level is such as $\Delta E_{\mathrm{min}} = 89.04$ cm$^{-1}$, and Mg I $9k-10l$, where $\Delta E_{\mathrm{min}}= 0.04$ cm$^{-1}$. The inelastic contribution of impact ions increases, and the contribution of the quadrupole term becomes negligible: It is less than $4\%$ for electrons and less than $1\%$ for ions.
\begin{figure}[!tH]
\vbox{
\centerline{\psfig{figure=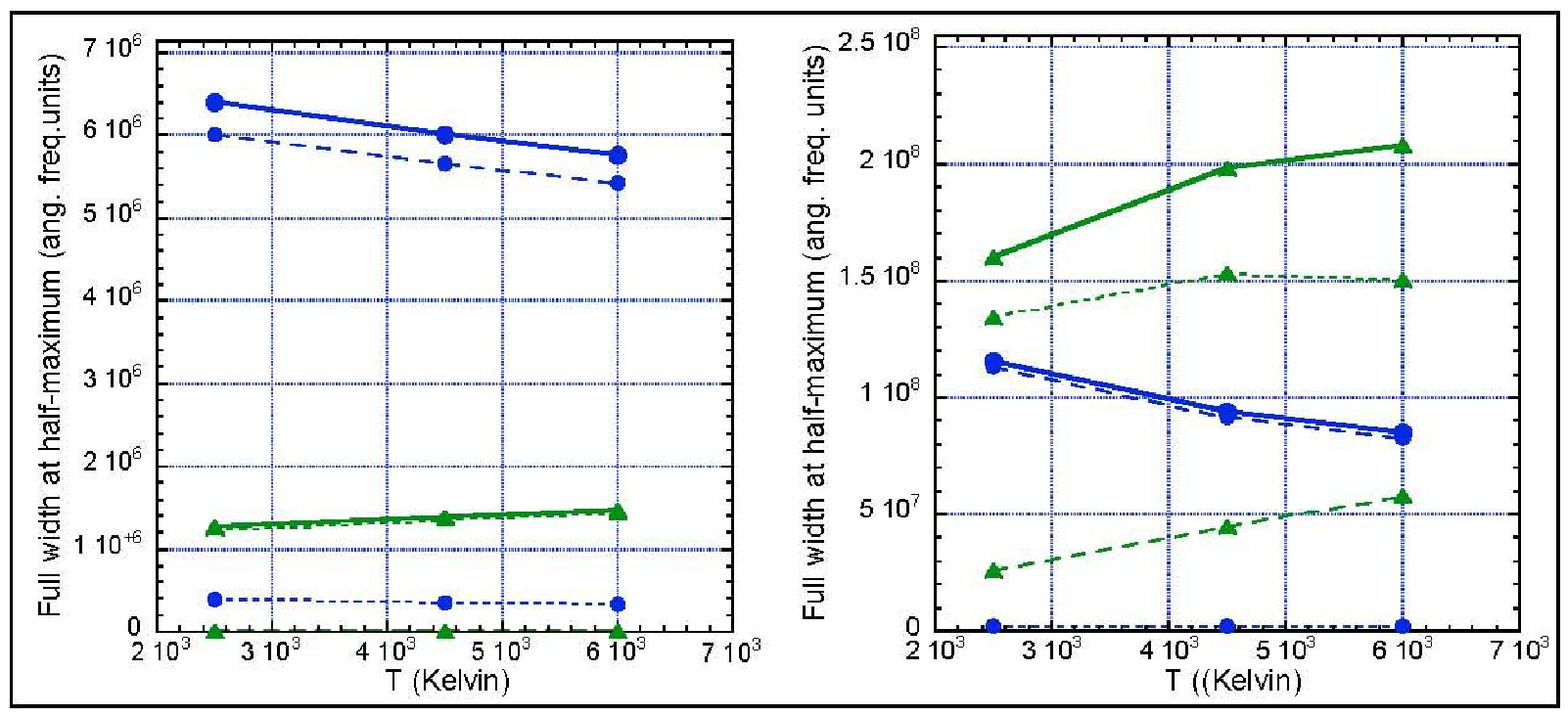,width=90mm,angle=0,clip=}}
\captionb{2}
{$w$ in angular frequency units as a function of the temperature $T$ in Kelvin, the electronic density is $10^{10}$ cm$^{-3}$. Full lines: total widths; dashed lines: inelastic contributions; dotted lines: elastic contributions; circles: electrons; triangles: Fe II ions. Left part: Mg I $5g^{1}G-6h^{1}H^{o}$.
Right part: Mg I $9k-10l$.
}}
\end{figure}

Then we consider the case of ion lines on the example of the Li-like Al XI ion (Dimitrijevi\'c \& Sahal-Br\'echot 1994 and STARK-B).
We begin with the case of the resonance line $2s-2p$, (Figure 3), $\Delta E_{\mathrm{min}}=1.8446$ $10^{6}$ cm$^{-1}$, and $\Delta E_{\mathrm{min}}/kT=2.6$ at $5$ $10^{5}$ K. Coulomb repulsion is high for colliding ions, and their contributions are less than $10 \%$. The quadrupole part of the elastic contribution is predominant. Proton contribution is higher than He III contribution at high temperatures, owing to the decrease of the Coulomb repulsion. For elastic electron collisions, the quadrupole part is dominant, but Feshbach resonances are important at low temperatures.
The case of  the $2p-5s$ line is completely different, because the $5s$ level is rather close to the $5p$ one: $\Delta E_{\mathrm{min}}/kT=1.39$ $10^{-2}$ at $5$ $10^{5}$ K. So the Coulomb repulsion is rather small for colliding ions, and the ionic width increases and becomes higher than the one due to electrons. The highest contribution is due to He III. This is due to the charge- and the reduced mass- effect.This is shown in Figure 4. The polarization potential prevails for elastic collisions because the involved levels are high.
\begin{figure}[!tH]
\vbox{
\centerline{\psfig{figure=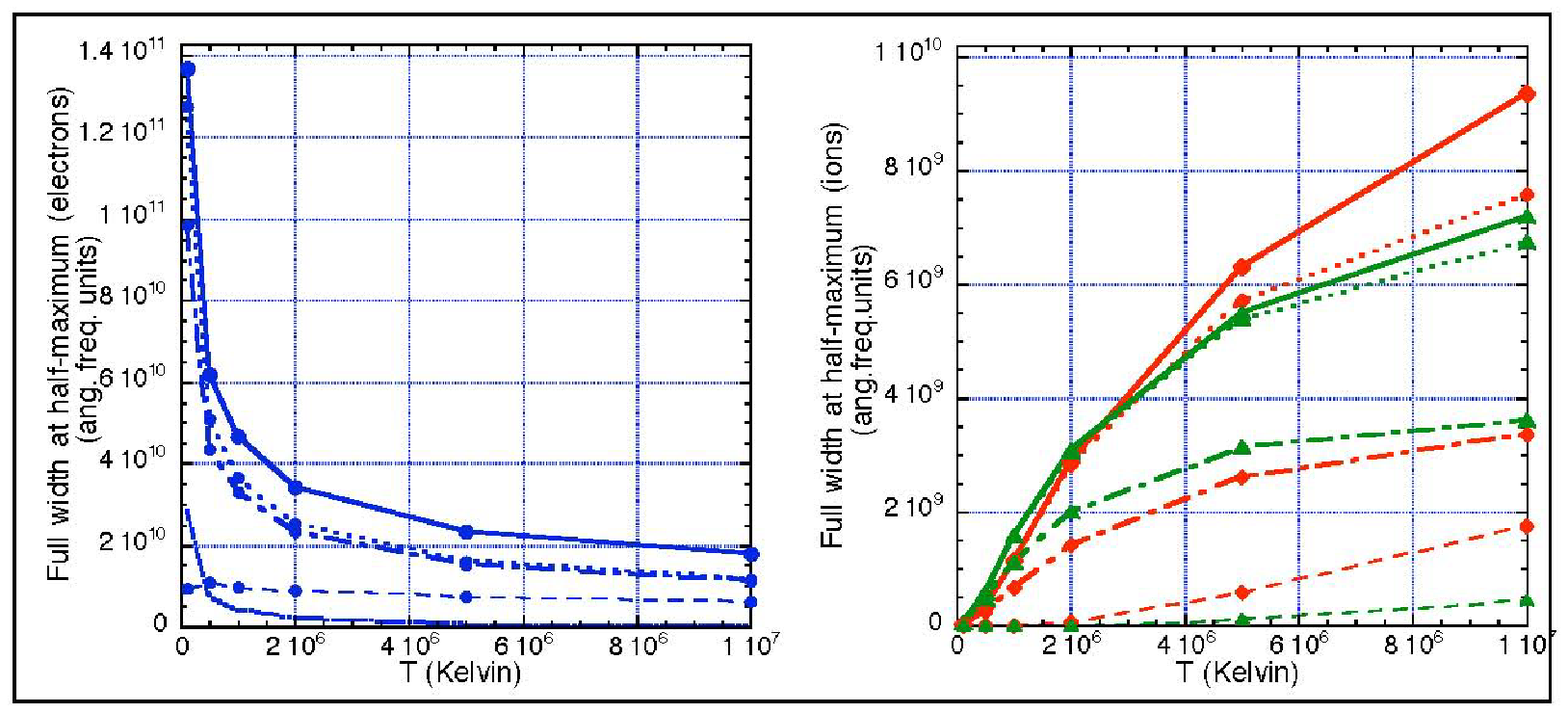,width=90mm,angle=0,clip=}}
\captionb{3}
{ Al XI  $2s^{2}S-2p^{2}P^{o}$: $w$ in angular frequency units.  $N = 10^{18}$ cm$^{-3}$. Full lines: total widths;  inelastic contributions: dashed lines; dotted lines: elastic contributions; dot-dashed lines: quadrupole contribution; dotted lines (small dots) without circles, Feshbach resonances;  circles: electrons; diamonds: protons; triangles: He III ions.
Left part: impact electrons.
Right part: impact ions.}}
\end{figure}
\begin{figure}[!tH]
\vbox{
\centerline{\psfig{figure=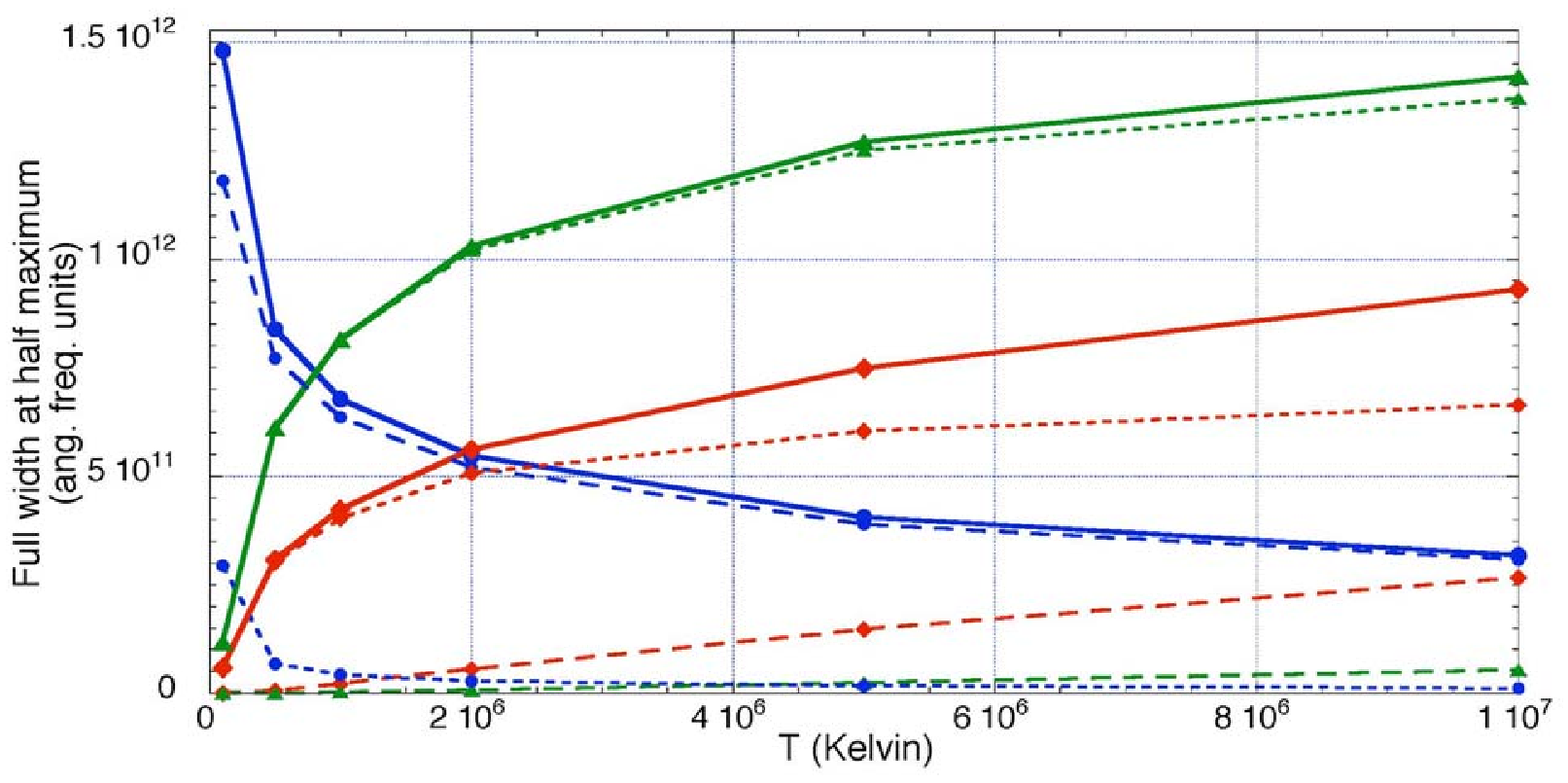,width=70mm,angle=0,clip=}}
\captionb{4}
{$w$ in angular frequency units: Al XI  $2p^{2}S-5s^{2}P^{o}$.   $N = 10^{18}$ cm$^{-3}$. Full lines: total widths;  dashed lines: inelastic contributions; dotted lines: elastic contributions;  circles: electrons; diamonds: protons;  triangles: He III ions.
}}
\end{figure}

This study explains why the width due to impact ions of Cr I lines ( Dimitrijevi\'c et al. 2005)  are higher than the widths due to electrons: there are  perturbing levels that are very close to the upper ones ($4.26$ and $14.14$ cm$^{-1}$). This abnormal situation is due to configuration interaction effects.

Concerning the behaviour with the charge of the perturber, the widths and the shifts increase linearly with $Z_{P}$ (Dimitrijevi\'c 1999 for instance among several papers) as expected by the SCP formulae.

 Second, we consider the effect of atomic structure and the charge of the radiating ion $Z_{A}$. First, it is expected than the width increases as $n^4$ when the principal quantum number $n$ increases. This is shown in Figure 5.
 \begin{figure}[!tH]
\vbox{
\centerline{\psfig{figure=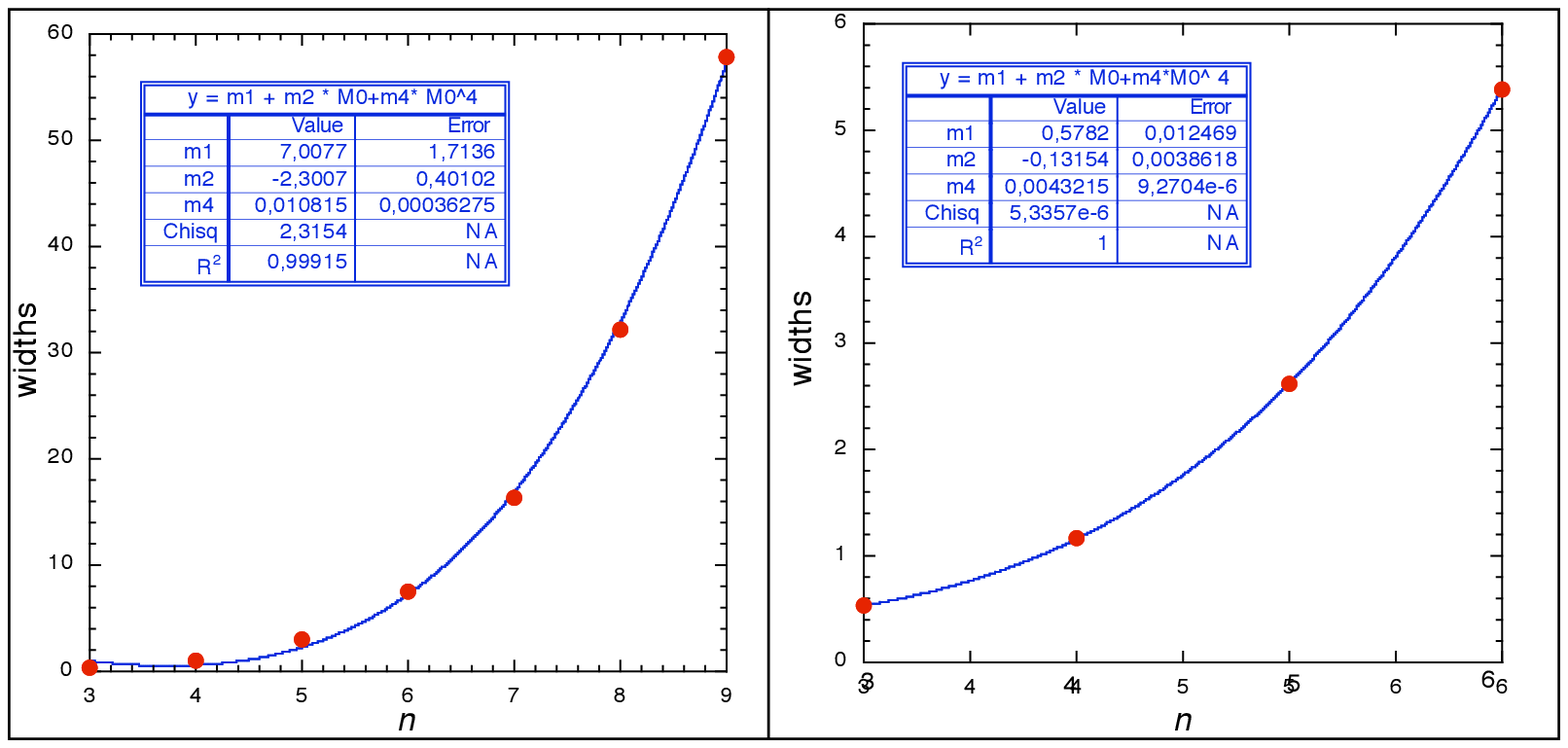,width=85mm,angle=0,clip=}}
\captionb{5}
{Behaviour of the principal quantum number $n$ of the widths (arbitrary units) of the series $3s-np$. Full circles: results of the calculations. Full line: least squares polynomial fitting (4th order). M0$=n$. The coefficients of the fitting, the $\chi^2$ and the correlation factor $R^2$ are given in the boxes.
Left: Na I. Right: Li II
}}
\end{figure}

The widths are predicted to vary as $Z_{eff}^{-2}$, whith $Z_{eff} = Z_{A}+1$. This is shown in Elabidi et al. (2009) for instance: cf. figure 20 of that paper, which shows a $-1.84$ slope for the $3s-3p$ transitions from C IV to P XIII .

Now we look at the influence of the chosen atomic structure for the SCP calculations. Larbi-Terzi et al. (2009) have shown on the example of the widths of the C II $3d-nf$ series  calculations that the differences in the results are very weak (less than 1\%) when the Coulomb approximation with quantum defect (Bates \& Damgaard 1949) oscillator strengths are used compared to the TOPBASE R-matrix calculations  (Cunto et al. 1973). In fact, C II is a simple atom and a simple atomic structure is sufficient. However, when highly charged ions or moderately charged ions like Si V (Ben Nessib et al. 2004) or Ne V (Hamdi et al. 2007)  are concerned, the choice of a good atomic structure becomes important. For these two ions, the difference can attain 25--30\% between the Bates \& Damgaard approximation and the more sophisticated method SUPERSTRUCTURE (Nussbaumer \& Storey 1978).

Finally, it will be pointed out that the behaviours of the fine structure widths of a multiplet  are not very sensitive to the fine structure splitting: for the $3s-3p$ multiplets of the Li-like series, the ratio of the widths of the two components only attains $1.12$ for P XIII (after Elabidi et al. 2009). This is quite negligible by looking at the accuracy of the calculations.

\sectionb{4}{FITTING FORMULAE AS FUNCTIONS OF TEMPERATURE}
\begin{figure}[!tH]
\vbox{
\centerline{\psfig{figure=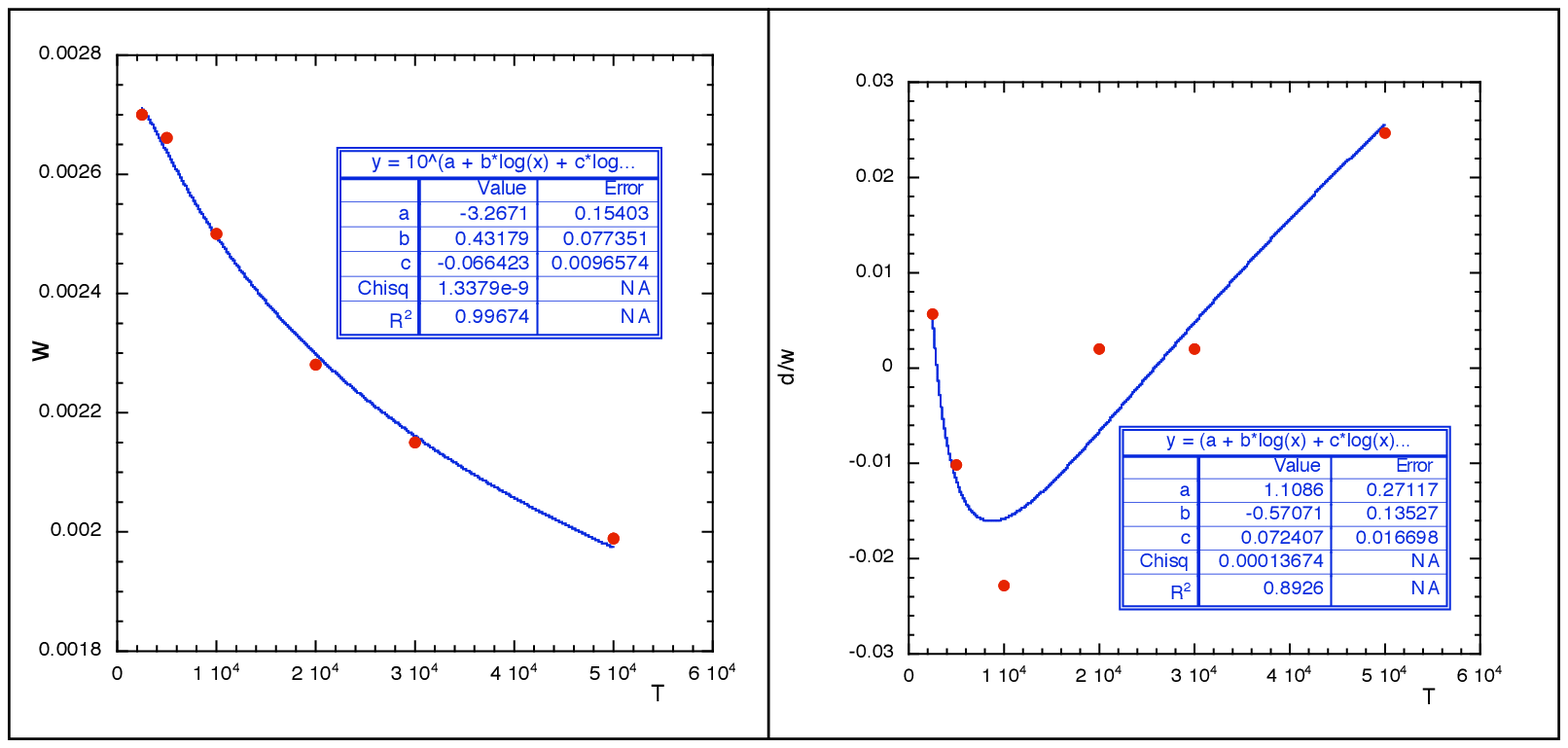,width=90mm,angle=0,clip=}}
\captionb{6}
{Example of fitting corresponding to Equation (1). Mg I $4f ^{1}F^{o}-6g ^{1}G$ at $10^{11}$ cm$^{-3}$  full circles: results of the SCP calculations, electron collisions; full line: fitting. The fitting coefficients (denoted as a, b, c in the figure),
$\chi^{2}$,
 and  $R^2$ are displayed in the boxes. $x$ is the temperature $T$ in Kelvin.
Left part: width.
Right part: shift/width
}}
\end{figure}
The theory and the SCP formulae show that the widths vary as $T^{-1/2}$ at low temperature and as $\mathrm{log}(T)/T^{1/2}$ at high temperatures. Among various papers, this was checked by Elabidi et al. (2009). But this is not sufficient for the users.
Astrophysics need fitting formulae and coefficients as functions of temperature for each line. In fact, such fitting coefficients are easier to  enter the modelling computing codes than  tables providing  widths and shifts for a set of temperatures.
So, for these astrophysical needs, we have obtained a simple but accurate fitting formula based on a least-squares method. It is logarithmic $+$ second degree polynomial:
\begin{equation}
\begin{array}{l}
 \log (w) = a_0  + a_1 \log (T) + a_2 \left[ {\log (T)} \right]^2, \\
 d/w = b_0  + b_1 \log (T) + b_2 \left[ {\log (T)} \right]^2. \\
 \end{array}
\end{equation}
Interestingly, this above $w$ fitting formula is to be compared to $w= C + A T^{B}$ which was proposed by Dimitrijevi\'c et al. (2007). The present one is more accurate, due to the second degree term of the expansion. However, none of them have a real physical sense.

Figure 6 shows an example of such a fitting: Mg I $4f ^{1}F^{o}-6g ^{1}G$ at $10^{11}$ cm$^{-3}$ (Dimitrijevi\'c \& Sahal-Br\'echot 1996, and other related papers cited in STARK-B). The fit is excellent for the width, but not so good for the shift: this is due to the fact that the accuracy for the calculation of the shift can be bad when it is very small. This bad shift example has been deliberately chosen for testing the accuracy of the fitting formula.

 The present coefficients will enter STARK-B in a near future under the form of complementary tables for each line.

\sectionb{5}{CONCLUSION}
We hope that the present study will help the users to interpret the results of SCP calculations, to obtain interpolated and extrapolated data that are not in the tables of STARK-B, and to enter the provided  fitting coefficients into their modelling codes for stellar atmospheres and star interiors.

\thanks{ A part of this work has been supported by VAMDC. VAMDC is funded under the ÒCombination of Collaborative Projects and Coordination and Support Actions  Funding Scheme of The Seventh Framework Program. Call topic: INFRA-2008-1.2.2 Scientific Data Infrastructure. Grant Agreement number: 239108. This work has also been supported by the cooperation agreement between Tunisia (DGRS) and France (CNRS) (project code 09/R 13.03, No.22637), by the Programme National de Physique Stellaire (INSU-CNRS), by the Paris Observatory and by the project 176002 of Ministry of Education and Science of Serbia.}

\References
\refb  Baranger M. 1958a, Phys. Rev. 111, 481

\refb  Baranger M. 1958b, Phys. Rev. 111, 494

\refb  Baranger M. 1958c, Phys. Rev. 112, 885

\refb  Bates D. R., Damgaard A. 1949, Phil. Trans. R. Soc. Lond. A, 242, 101

\refb  Ben Nessib N., Dimitrijevi\'c M.S., Sahal-Br\'echot S.,2004, A\&A, 423, 397



\refb Cunto W., Mendoza C.,Ochsenbein F.,Zeippen C.J.1993, A\&A, 275, L5

\refb Dimitrijevi\'c M.S., Sahal-Br\'echot S. 1984, JQSRT, 31, 301

\refb Dimitrijevi\'c M.S., Sahal-Br\'echot S. 1994, A\&AS, 105, 245 

\refb Dimitrijevi\'c M.S., Sahal-Br\'echot S. 1996, A\&AS, 117, 127 

\refb Dimitrijevi\'c  M.S. 1999, SAJ, 159, 65

\refb Dimitrijevi\'c M.S., Ryabchikova T., Popovi\'c L.\v C. et al. 2005, A\&A, 435, 1191

\refb Dimitrijevi\'c M.S., Ryabchikova T., Simi\'c Z. et al. 2007, A\&A, 469, 681





\refb  Dubernet M.L., Boudon V., Culhane J.L. et al. 2010, JQSRT, 111, 2151;  http://www.vamdc.eu


\refb  Elabidi H., Ben Nessib N.,  Sahal-Br\'echot S. 2009, EPJD, 54, 51

\refb  Fleurier C., Sahal-Br\'echot S., Chapelle J. 1977, JQSRT,  17, 5954

\refb 	Hamdi R., Ben Nessib N., Dimitrijevi\'c M.S., Sahal-BrŽchot S. 2007, ApJS, 170, 243

\refb  Larbi-Terzi N., Sahal-Br\'echot S., Ben Nessib N. et al. 2010: AIP Conf. Proc., 1273, 428 

\refb Nussbaumer H., Storey J. P.,1978, A\&A, 64, 139

\refb Mahmoudi W., Ben Nessib N., Sahal-Br\'echot S. 2008, EPJD, 47, 7

\refb Popovi\'c  L. $\check {\rm C}$, Dimitrijevi\'c M. S., Simi\'c Z. et al.  2008, New Astron. 13 85

\refb  Rixon G.,  Dubernet M. L., Piskunov N. et al. 2010, 7th International Conference on Atomic and Molecular Data and their Applications  - ICAMDATA-2010, Vilnius, (Lithuania), 21-24 September 2010 ; AIP Conf. Proc.,  1344, 107

\refb  Sahal-Br\'echot S.1969a, A\&A , 1, 91

\refb  Sahal-Br\'echot S. 1969b, A\&A , 2, 322

\refb  Sahal-Br\'echot S. 1974, A\&A, 35, 319

\refb  Sahal-Br\'echot S. 2010, J. Phys.: Conf. Ser., 257, 012028

\refb Sahal-Br\'echot S., Dimitrijevi\'c M.S., Moreau N. 2008, http://stark-b.obspm.fr

\refb  Stehl\'e C., and Feautrier N. 1985, J.Phys.B: At. Mol. Phys., L43

\end{document}